\documentclass{article}
\usepackage{amsmath}
\usepackage{amssymb}
\usepackage{amsthm}
\usepackage{graphicx}
\usepackage{psfrag}
\usepackage[hang, nooneline]{subfigure}

\author{Natascha Riahi \footnote{e-mail address: natascha.riahi@gmx.at}
\\ University of Vienna, Faculty of Physics, Gravitational Physics
\\ Boltzmanng. 5, 
1090 Vienna, Austria}
\date{}
\title{How to use Unimodular Quantum Cosmology for the Prediction of a late-time Classical Universe ? }

\begin{document}

\maketitle

\begin{abstract}
Unimodular quantum cosmology admits wavepacket solutions that evolve according to a
 kind of Schr\"odinger equation. Though this theory is equivalent to general relativity on
 the classical level, its canonical structure is different and the problem of time 
 does not occur. 
We present an Ehrenfest theorem for the long term evolution of the expectation value of the scale factor
for a spatially flat Friedmann universe with a scalar field. We find that the classical and the quantum behaviour in 
the asymptotic future coincide for the special case of a 
massless scalar field. We examine the general behaviour of uncertainties in order to single out models that can lead to a classical universe. 
\end{abstract}

\section{Introduction}
The canonical quantization of general relativity leads to the so-called  problem of time (see \cite{Kiefer} and references therein).
In most non-perturbative approaches of quantum gravity time has disappeared from the theory and is seen as an artifact of the classical limit.
Here we investigate quantum cosmology in the  framework of unimodular gravity.  This theory is practically equivalent to 
general relativity at the classical level, but since it has a different canonical structure
time does not disappear from the quantum theory (\cite{HT}) and it is possible to study the evolution of wave packet solutions and the behaviour of 
expectation values compared to the classical evolution of their counter parts.

In \cite{Uni1} we constructed a class of unitarily evolving solutions  
with a negative expectation value of the Hamiltonian for the special case of a spatially flat universe with a massless scalar field. 
Investigating a special example, we found that the classical and quantum dynamics of the scale factor coincide for the asymptotic future
(though with significant spread).
Here we compare the expectation value of the scale factor to the  evolution of its classical counterpart for solutions of
the spatially flat Friedmann universe with an arbitrary scalar field which yields Ehrenfest theorem for the late time behaviour.
We examine the evolution of the uncertainties  in order to single out models that can lead to a classical universe in the asymptotic future.
\section{About Unimodular Gravity}

We start  with the Einstein Hilbert action (\ref{EH}) 
\begin{equation}
\label{EH}
S_{EH}=\frac{1}{2 \kappa}\int _{\mathcal{M}}d^4x\,\sqrt{-g}\,(R-2 \Lambda)-
\frac{1}{ \kappa} \int _{\mathcal{\partial M}}d^3x\,\sqrt{h}\,K\,,
\end{equation}
where 
\[
 \kappa=\frac{8 \pi G}{c^4}
 \]
contains the velocity of light $c$ and the gravitational constant $G$. We also take into account the matter action $S_{m}$ that describes the fields.
If we vary the action $S=S_{m}+S_{EH}$  with respect to the metric $g_{\mu \nu }$ under the restriction $-g=1$, we obtain Einsteins equations with an arbitrary additional
constant $\Lambda$ , that can be identified with the cosmological constant of general relativity (\cite{HT}). 
\begin{subequations}
\label{Unimod}
\begin{align}
& R_{\mu \nu}-\frac{1}{2}g_{\mu\nu}R=\kappa\, T_{\mu \nu}-\Lambda \,g_{\mu \nu } \label{Unimoda}\\
& \sqrt{-g}-1=0 \label{Det} \,.
\end{align}
\end{subequations}
This theory is called unimodular gravity.
Any solution of unimodular gravity (\ref{Unimod}) is also a solution of general relativity 
for a specific cosmological constant and vice versa. The only difference between the two theories is, that $\Lambda$ is
a natural constant in general relativity while it is a conserved quantity in unimodular gravity. But since
in both theories the cosmological constant can not vary over the whole universe, we would have to investigate
different universes to determine if solutions with different $\Lambda$ exist (unimodular theory) or if
$\Lambda$ is a ''true'' natural constant.  So the two theories are practically indistinguishable.
Nevertheless the canonical structure of the theories differs (\cite{HT})and therefor the quantization of unimodular theory 
yields different results compared to the quantization of general relativity (\cite{Unruh}).

\section{The Unimodular Hamiltonian of a spatially flat Friedmann Universe}
The metric of a homogeneous and isotropic spacetime (Friedmann universe)

\begin{equation}
\label{MFriedmann}
  ds^2=-N^2(t) c^2 dt^2+a^2(t) d\Omega^2_{3} 
  \end{equation}
  is characterized by the lapse function N(t) and the scale factor $a(t)$. If the spatial curvature is zero,
$d\Omega^2_{3}$ is the line element of three-dimensional flat space.

Inserting the metric into the Einstein-Hilbert action (\ref{EH}) with $\Lambda=0$ yields (\cite{Kiefer})
\begin{align}
& S_{EH}=\frac{3}{ \kappa}\int dt\,N\, \left(-\frac{\dot{a}^2 a}{c^2 N^2}\right)\,v_{0}\,,
\nonumber\,
\end{align}
where $v_{0}$ is is the  volume of the spacelike slices according to (\ref{MFriedmann}).

The action of a scalar field in a Friedmann universe  (\ref{MFriedmann}) reads
\begin{equation}
\label{Sm}
S_{m}=\int\, dt\, N a^3 \left(\frac{\dot{\phi}^2}{2 N^2 c^2}-V(\phi) \right) v_{0}\,.
\end{equation}

Using the unimodular condition for the lapse function 
$N=a^{-3}$, we find for the Hamiltonian (\cite{Uni1})

of the unimodular theory

\begin{equation}
\label{Huni}
 H_{uni}=\frac{c^2}{2}\frac{p_{\phi}^2}{a^6}-\frac{c^2}{4 \epsilon}\frac{p_{a}^2}{a^4}
 + V(\phi)
 \end{equation}
The Hamiltonian is a conserved quantity and not a constraint as in general relativity

The canonical quantization of this  Hamiltonian yields

\begin{equation}
 \label{QImplulse}
 \hat{p}_{a}=-i \hbar \frac{\partial}{\partial a}\,,\quad
  \hat{p}_{\phi}=-i \hbar \frac{\partial}{\partial \phi}\,,
\end{equation}

\begin{equation}
\label{Huv}
\widehat{H}=\frac{\hbar^2 c^2}{4 \epsilon}\frac{1}{a^5}\frac{\partial}{\partial a} a \frac{\partial}{\partial a}
 -\frac{\hbar^2 c^2}{2} \frac{1}{a^6 }\frac{\partial^2}{\partial \phi^2}+V(\phi)\,.
\end{equation}

Here we have chosen the factor ordering that gives the part of the Hamiltonian that is quadratic in the momenta the form of a 
Laplace Beltrami 
operator (\cite{Kiefer}).

The evolution of the wavefunction  $\psi(a,\phi,t)$ is  determined by
\begin{equation}
\label{S}
\widehat{H}\psi=i \hbar \frac{\partial}{\partial t} \psi\,.
 \end{equation}
 The Hamiltonian is symmetric with respect to the inner product defined by the measure $a^5 da d\phi$, where 
 $a \in (0,\infty)$ and $\phi \in (-\infty,\infty)$.

Applying the coordinate transformations

\begin{subequations}

\label{Trafo}

\begin{equation}
\label{Trafo1}
 A=a^3/3 \qquad B= \frac{3}{\sqrt{2 \epsilon}}\phi\,,
\end{equation}

and 
\begin{equation}
 \label{Trafo2}
u=A e^{-B}\qquad v=A e^{B}\,.
\end{equation}

\end{subequations}

We obtain the Hamilton operator

\begin{equation}
\label{Hop}
 \widehat{H}=\frac{\hbar^2 c^2}{\epsilon}\frac{\partial^2}{\partial u \partial v }+V\left(\frac{u}{v}\right)
\end{equation}

The volume element is given by
 $du dv$ and $u \in (0,\infty)$,
$v \in (0,\infty)$. 

The classical unimodular Hamiltonian then reads

\[
 H=-\frac{c^2}{\epsilon}p_{u}\,p_{v}+V\left(\frac{u}{v}\right)\,.
 \]

 The Laplace-Beltrami factor ordering ensures that the quantizations of the Hamiltonian commutes with
 coordinate transformation if we understand classical transformations as 
 canonical point transformations (see also cite {KucharP2})
 
\section{General properties of the time evolution}

In \cite{Uni1} we have derived a class of unitarily evolving wavepacket solutions of (\ref{S}) for the special case of a
massless scalar field ($V=0$). We found that these solutions fulfill in the late phase of time evolution
\begin{equation}
\label{OLimes}
 \lim_{t \rightarrow \infty}\, \psi(0,v,t)=
  \lim_{t \rightarrow \infty}\,
  \psi(u,0,t)=0\,\,.
\end{equation}

Now we  assume that a wavepacket solution for a general scalar field can be found and that it also fulfills (\ref{OLimes}).
We investigate the physical behaviour of these solutions compared to the evolution of the classical quantities.

As in ordinary quantum mechanics, the evolution of the expectation values of an observable $\widehat{O}$ with respect to a 
solution $\psi(u,v,t)$ of the Schr\"odinger equation (\ref{S}) is given by

\begin{equation}
 \frac{d}{dt}\langle\psi| \widehat{O} |\psi \rangle=
 -\frac{i}{\hbar}
 \left(\langle \psi|\widehat{O}\widehat{H}|\psi \rangle
 -\langle\widehat{H} \psi|\widehat{O}\psi \rangle\right)\,.
\end{equation}

This implies the equation
\begin{equation}
\label{HGleichung}
 \frac{d}{dt}\langle\psi| \widehat{O} |\psi \rangle= -\frac{i}{\hbar}
\left \langle \psi\left[\widehat{O},\widehat{H}\right]\psi\right\rangle \,,
\end{equation}
only if  $\widehat{O} \psi$  obeys

\begin{equation}
 \langle \widehat{H}\psi|\widehat{O}\psi \rangle\stackrel{!}{=}
 \langle \psi|\widehat{H}\widehat{O}|\psi \rangle\,.
\end{equation}
In the case of the Hamiltonian(\ref{Hop}), this condition reads
\begin{equation}
\label{Bedingung0}
\int_{0}^{\infty}\psi^{*}(u,v,t)\frac{\partial}{\partial v}\widehat{O}\psi(u,v,t)dv
\mathop {\Big |_{u=0}}-
\int_{0}^{\infty}\left(\frac{\partial}{\partial u}\psi^{*}(u,v,t)\right)\widehat{O}\psi(u,v,t)du\mathop {\Big |_{v=0}} =0\,.
\end{equation}
We find that this condition is fulfilled in the limit
$t \rightarrow \infty$ for wavepackets with the property (\ref{OLimes}). This means that the 
time evolution that we derive by the application of (\ref{HGleichung}), gives the correct
result for the limit $t \rightarrow \infty$, and is approximately valid in the late phase of time evolution. 

We find for the variable $u v$, related to the scalefactor by
\[
 A^2=u v=\frac{a^6}{9}\,,
\]
\begin{subequations}
\label{Ehrenfest}
\begin{align}
\label{Ehrenfest1}
   \lim_{t \rightarrow \infty} \frac{d^2}{dt^2}\langle A^2 \rangle=\frac{c^2}{\epsilon}
  \left\langle -2\,\widehat{H} + V +
  u \frac{\partial V}{\partial u}+ v \frac{\partial V}{\partial v }\right\rangle \,,
  \end{align}
  whereas the classical time evolution reads
  \begin{align}
  \label{Ehrenfest2}
  \lim_{t \rightarrow \infty} \frac{d^2}{dt^2}A^2 =\frac{c^2}{\epsilon}\left(
-2 \,H + V +
   u \frac{\partial V}{\partial u}+ v \frac{\partial V}{\partial v }\right)\,.
  \end{align}
  \end{subequations}
  So we see that in the special case of a massless scalar field the late time behaviour of the classical scalefactor and its 
  expectation value according to unimodular quantum cosmology are the same. In general (\ref{Ehrenfest}) represents the Ehrenfest theorem for
  unimodular quantum cosmology.
  
  The result for the uncertainty
$\Delta( A^2)$ in the special case of a massless scalar field reads

\[
\lim_{t\rightarrow \infty}\frac{d^4}{dt^4}(\Delta( A^2))^2=
\frac{24 c^2}{\epsilon^2} (\Delta H )^2 \,,
\]
which implies that the uncertainty $\Delta( A^2)$ is monotonically 
growing with $t^2$ in the late phase of time evolution. 

\section{Evolution of uncertainties for general matter models}

Since we are interested in the evolution of unvertainties for late times of the universe we will 
from now on  assume that (\ref{Bedingung0})
is fulfilled in good approximation for late times, so that we can use (\ref{HGleichung})and so 
all dynamical equations we derive are valid in the asymptotic future ($\mbox{limit}\,t\rightarrow \infty$).
We find for the dynamics of the variables $u,v$ and the associated momenta $\widehat{p_{v}}, \widehat{p_{u}}$
\begin{subequations}
\label{DynObs}
\begin{align}
\label{DynObs1}
&\frac{d}{dt}\langle v \rangle=-\mu \left \langle \widehat{p_{u}} \right \rangle\,,&
&\frac{d}{dt}\langle u \rangle=-\mu \left \langle \widehat{p_{v}}\right  \rangle\,,
&\\
\label{DynObs2}
&\frac{d}{dt}\langle \widehat{p_{u} }\rangle=-\left \langle \frac{\partial{V}}{{\partial v}} \right \rangle\,,&
&\frac{d}{dt}\langle \widehat{p_{v} }\rangle=-\left \langle \frac{\partial{V}}{{\partial u}} \right \rangle\,,\,
&
\end{align}

\end{subequations}

where $\mu=c^2/\epsilon=2 \pi G/(3 c^2) $.

If $\frac{\partial{V}}{{\partial v}},\frac{ \partial{V}}{ {\partial u}}$ fulfill
\begin{equation}
\label{Deviation}
 \langle F(u,v) \rangle \approx F(\langle u\rangle, \langle v \rangle)
\end{equation}

the equations constitute a closed system of ordinary differential equations for the expecation values, which coincide with the 
classical evolution equations according to (\ref{Huv}).

The four observables $\hat{u},\hat{v},\widehat{p_{v}}, \widehat{p_{u}}$ give rise to ten uncertainties 
$\Delta u^2, \Delta v^2$, $(\Delta p_{u})^2, \Delta (p_{v})^2$,
$\Delta (u, v), \Delta (u, p_{u}), \,\Delta (v, p_{v}),\, \Delta (u, p_{v}),\, \Delta (v, p_{u}),\, \Delta ( p_{u},p_{v})$,
where the mixed uncertainties of two observables are defined by

\[
 \Delta(\widehat{v_{1}},\widehat{v_{2}})=
 \frac{1}{2}\left \langle \widehat{v_{1}}\widehat{v_{2}}+\widehat{v_{2}}\widehat{v_{1}}
 -2 \left\langle \widehat{v_{1}}\right\rangle \left\langle\widehat{v_{1}}\right\rangle
 \right \rangle
\]
The application of (\ref{HGleichung}) yields the evolution equations (where we only denote the first three here):
\begin{subequations}
\label{Evolution1}
\begin{align}
&\frac{d}{dt} (\Delta u)^2 =-2 \mu  (\Delta(u,p_{v}) 
\\
& \frac{d}{dt} (\Delta v)^2  =-2 \mu  (\Delta(v,p_{u}))
\\
& \frac{d}{dt}\Delta(u,p_{v})=-\mu \Delta(p_{v})^2
-\left\langle \frac{\partial{V}}{\partial{v}}u\right\rangle
+\left\langle \frac{\partial{V}}{\partial{v}}\right\rangle\left\langle u \right\rangle
\\
\nonumber
&...
\end{align}
\end{subequations}
Expanding the occuring derivatives of $V$ around the expectation values given by (\ref{DynObs1}) and neglecting all
stochastic moments of higher than second order yields a closed system of linear differential equations for the variances 
(quadratic uncertainties)
\begin{subequations}
 \label{Evolution2}

\begin{align}
&\frac{d}{dt} (\Delta u)^2 =-2 \mu  (\Delta(u,p_{v}) 
\\
& \frac{d}{dt} (\Delta v)^2  =-2 \mu  (\Delta(v,p_{u}))
\\
& \frac{d}{dt}\Delta(u,p_{v})=-\mu \Delta(p_{v})^2-V_{22}\Delta(u,v)-V_{12}(\Delta u)^2
\\
& \frac{d}{dt}\Delta(v,p_{u})=-\mu \Delta(p_{u})^2-V_{11}\Delta(u,v)-V_{12}(\Delta v)^2
 \\
 &\frac{d}{dt}\Delta(u,v)=-\mu \Delta(v,p_{v})-\mu \Delta(u,p_{u})
 \\
 &\frac{d}{dt} (p_{u})^2 =-2     V_{11} (\Delta(u,p_{u}) -2 V_{12} (\Delta(v,p_{u}) 
\\
&\frac{d}{dt} (p_{v})^2 =-2     V_{22} (\Delta(v,p_{v}) -2 V_{12} (\Delta(u,p_{v}) 
\\
& \frac{d}{dt}\Delta(p_{u},p_{v})=-V_{11}\Delta(u,p_{v})-V_{12}\Delta(p_{v},v)
-V_{22}\Delta(v,p_{u})-V_{12}\Delta(p_{u},u)
\\
& \frac{d}{dt}\Delta(u,p_{u})=-\mu \Delta(p_{u},p_{v})-V_{11}(\Delta u)^2-V_{12}\Delta(u, v)
\\
& \frac{d}{dt}\Delta(v,p_{v})=-\mu \Delta(p_{u},p_{v})-V_{22}(\Delta v)^2-V_{12}\Delta(u, v)\,,
\end{align}

\end{subequations}
where we have chosen the abbreviations
\[
 V_{11}=\frac{\partial^2 V}{\partial v^2 }
 \,,\quad V_{22}=\frac{\partial^2 V}{\partial u^2 }
 \,
 \quad
 V_{12}=\frac{\partial^2 V}{\partial v \partial u}\,.
\]
Given that the wavefunction ensures that all higher moments are smaller than the variances all uncertainies remain
bounded if the system (\ref{Evolution2}) is stable. 

This is especially the case if (see f.i.\cite{Coppel})
$V_{11},V_{22},V_{12}$ converge to fixed values $v_{11},v_{22},v_{12}$ for $t\rightarrow \infty$
 and fulfill 
\begin{align}
 & \left|V_{12}\right|\,>\,\sqrt{V_{11}V_{22}},\quad  V_{12}<0,\, V_{11}\cdot V_{22} >0, \mbox{for} \,t>t_{0}\,, \\
 & \left|v_{12}\right|\,>\,\sqrt{v_{11}v_{22}},\quad  v_{12}<0,\, v_{11}\cdot v_{22} >0
 \end{align}
  since then all roots of (\ref{Evolution2}) have zero real part and are of simple type, and if the matrix 
  $\mathbf{A(t)}$,characterizing (\ref{Evolution2}),
fulfills
  \[
  \int \limits_{t_{0}}^{\infty} \left|\mathbf{A'(t)} \right| dt\,<\,\infty\,.
  \]

Moreover, if we neglect deviations from the classical dynamics which would be of the order of magnitudes of uncertainties
(\ref{Deviation}) we can always evaluate the solutions (\ref{Evolution2}) inserting the classical solutions for
$\left\langle u\right\rangle,\left\langle v\right\rangle,
\left\langle \widehat{p_{u}}\right\rangle,\left\langle \widehat{p_{v}}\right\rangle$
and verify if the initial assumption of small uncertainties is justified.


\begin{thebibliography}{99}
\bibitem{Kiefer}Kiefer,C.: Quantum gravity, Oxford university press (2012)
\bibitem{HT}  Henneaux,M. , Teitelboim,C. (1989): 
 The cosmological constant and general covariance, Phys.Lett.B  222/2, 195
 \bibitem{Uni1} N.Riahi (2017): Wavepacket evolution in unimodular quantum cosmology, Galaxies 2018,6(1),8
  \bibitem{Unruh}  Unruh,W.G. (1989): Unimodular theory of canonical quantum gravity, Phys.Rev.D 40/4, 1048-051
% \bibitem{KucharP1} K.Kuchar (1986): Hamiltonian dynamics of gauge systems, Phys. Rev. D 34/10 3031-3043
\bibitem{KucharP2} K.Kuchar (1986): Hamiltonian dynamics of gauge systems, Phys. Rev. D 34/10 3043-3057
\bibitem{Coppel}
Coppel,W.A.: Stability and Asymptotic Behaviour of Differential Equations, Heath and Company, Boston(1965),
p. 113
%\bibitem{DLS} Daughton,A ,  Louko,J.  Sorkin,R.D. (1993): Initial conditions and unitarity
%in unimodular quantum cosmology, arxiv: gr-qc/9305016 
%\bibitem{Chiou}Chiou, D. , Geiller,M. (2010): Unimodular loop quantum cosmology, Phys. Rev. D 82, 064012
% \bibitem{Masden} Masden,M.S. (1985): A note on the equation of state of a scalar field , Astrophysics
% and Space Science 113/205-207

 %\bibitem{Za} Zayed,A.: Handbook of Function and Generalized Function Transformations, CRC Press 1996 
%\bibitem{Erdely} Erdely,A.: Tables of Integral Transforms, McGraw-Hill Book Company, 1954
%\bibitem{Magnus} Magnus,W., F.Oberhettinger, R.P.Soni: Formulas and Theorems for the 
 % Special Functions of Mathematical Physics, Springer-Verlag 
 % \bibitem{Ro}  Robinett,R.(2004): Quantum wave packet revivals, Phys.Rep. 251
 % \bibitem{CS} Craig,C.A., Singh,P. (2010): Consistent probabilities in Wheeler-De Witt quantum
 % cosmology, Phy.Rev.D 82/12, 123526 
  
\end{thebibliography}
\end{document}